\documentclass{iopart}
\usepackage{graphicx}
\usepackage{hyperref}
\usepackage{url}
\usepackage{color}
\newcommand{\eqref}[1]{(\ref{#1})}
\usepackage{Tomspeak}
\renewcommand{\tw}[1]{#1}
\newcommand{\rr}{\mathbf{r}}
\newcommand{\drr}{\delta \rr}
\newcommand{\mmin}{_{\rm min}}
\newcommand{\mmax}{_{\rm max}}
\newcommand{\linprog}{\texttt{linprog} }
\newcommand{\CC} {\mathtt{C}}
\begin{document}

\title{Signal Transmissibility in Marginal Granular Materials}

\author{Matthew B. Pinson and Thomas A. Witten \footnote{corresponding author: email t-witten@uchicago.edu} }
\address{James Franck Institute, University of Chicago, 929 East 57th Street, Chicago, Illinois 60637, USA \par}

\maketitle

\begin{abstract}

We examine the ``transmissibility" of  a simulated two-dimensional pack of frictionless disks formed by confining dilute disks in a shrinking, periodic box to the point of mechanical stability.  Two opposite boundaries are then removed, thus allowing a set of free motions.  Small free displacements on one boundary then induce proportional displacements on the opposite boundary.  Transmissibility is the ability to distinguish different perturbations by their distant responses.  We assess transmissibility by successively identifying free orthonormal modes of motion that have the {\em smallest} distant responses.  The last modes to be identified in this ``pessimistic" basis are the most transmissive.  The transmitted amplitudes of these most transmissive modes fall off exponentially with mode number.  Similar exponential falloff is seen in a simple elastic medium, though the responsible modes differ greatly in structure in the two systems.  Thus the marginal pack's transmissibility is qualitatively similar to that of a simple elastic medium.  We compare our results with recent findings based on the projection of the space of free motion onto interior sites.  
\end{abstract}
\submitto{\JPCM}
\section{Introduction}
\tw{Granular materials have aroused great current interest as realizations of solidification via confinement.  Idealized granular systems of packed, frictionless spheres show a well-characterized, critical ``jamming transition"\cite{Silbert2005}.  This transition is believed to be relevant for frictionless packed materials such as dense colloidal suspensions\cite{Duri:2009yq}.  The jamming transition is also of interest for glass-forming liquids and real granular materials.  The sharpness of the transition\cite{vanHecke2010}} has been utilised in applications such as the jamming gripper \cite{Brown2010}. Another important characteristic, in marginally-jammed granular materials \cite{Silbert2005} as in other critical materials, is a diverging length scale as the critical point is approached. For example, when a single constraint is removed from a marginally-jammed system of particles (by removing from a particular pair of particles the condition that they may not overlap), the resulting zero-energy mode has been shown to have typical size set by the size of the system \cite{Wyart2005,Lerner2013,Muller2015}.

\begin{figure}
\includegraphics[width=0.5\textwidth]{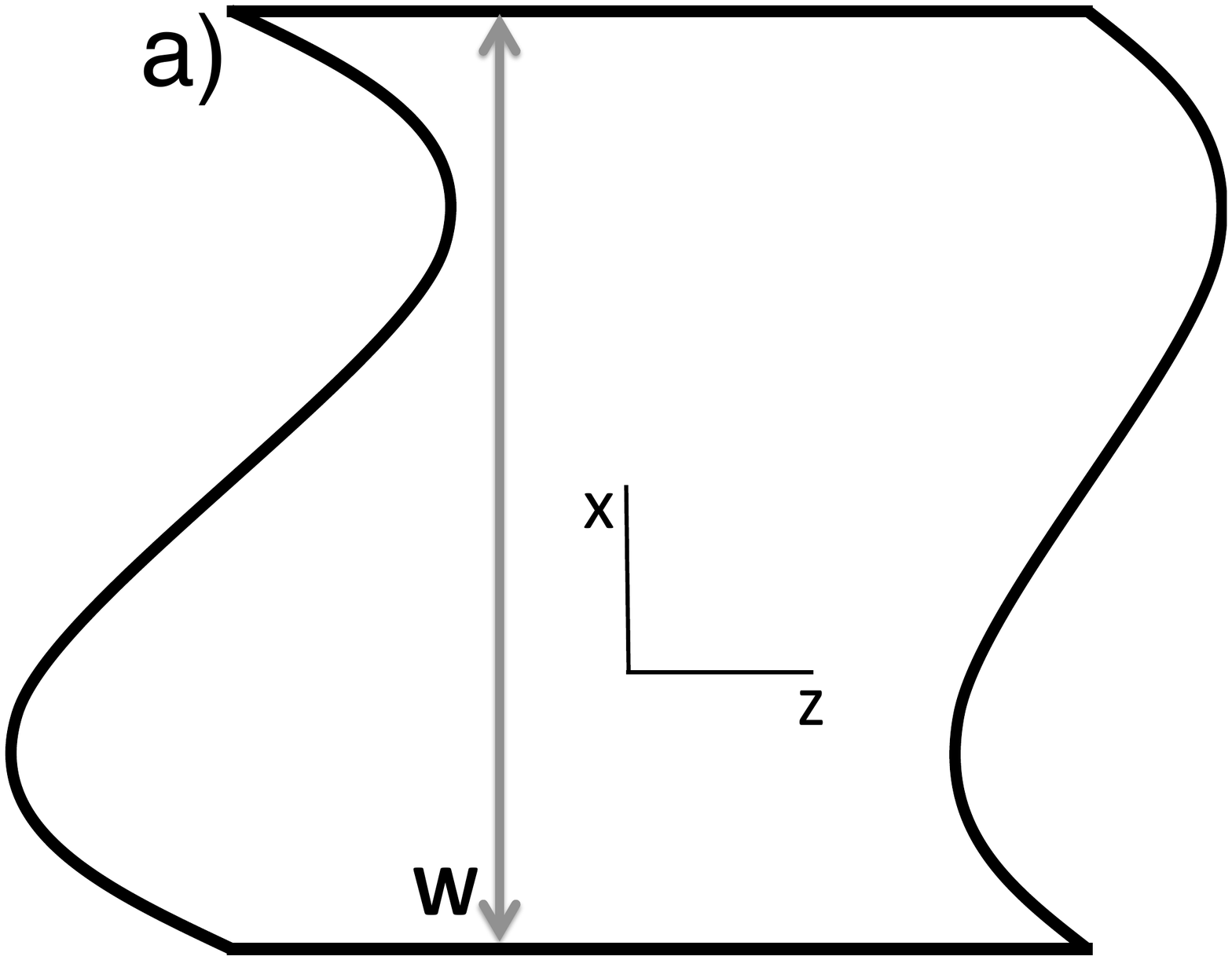}
\includegraphics[width=0.5\textwidth]{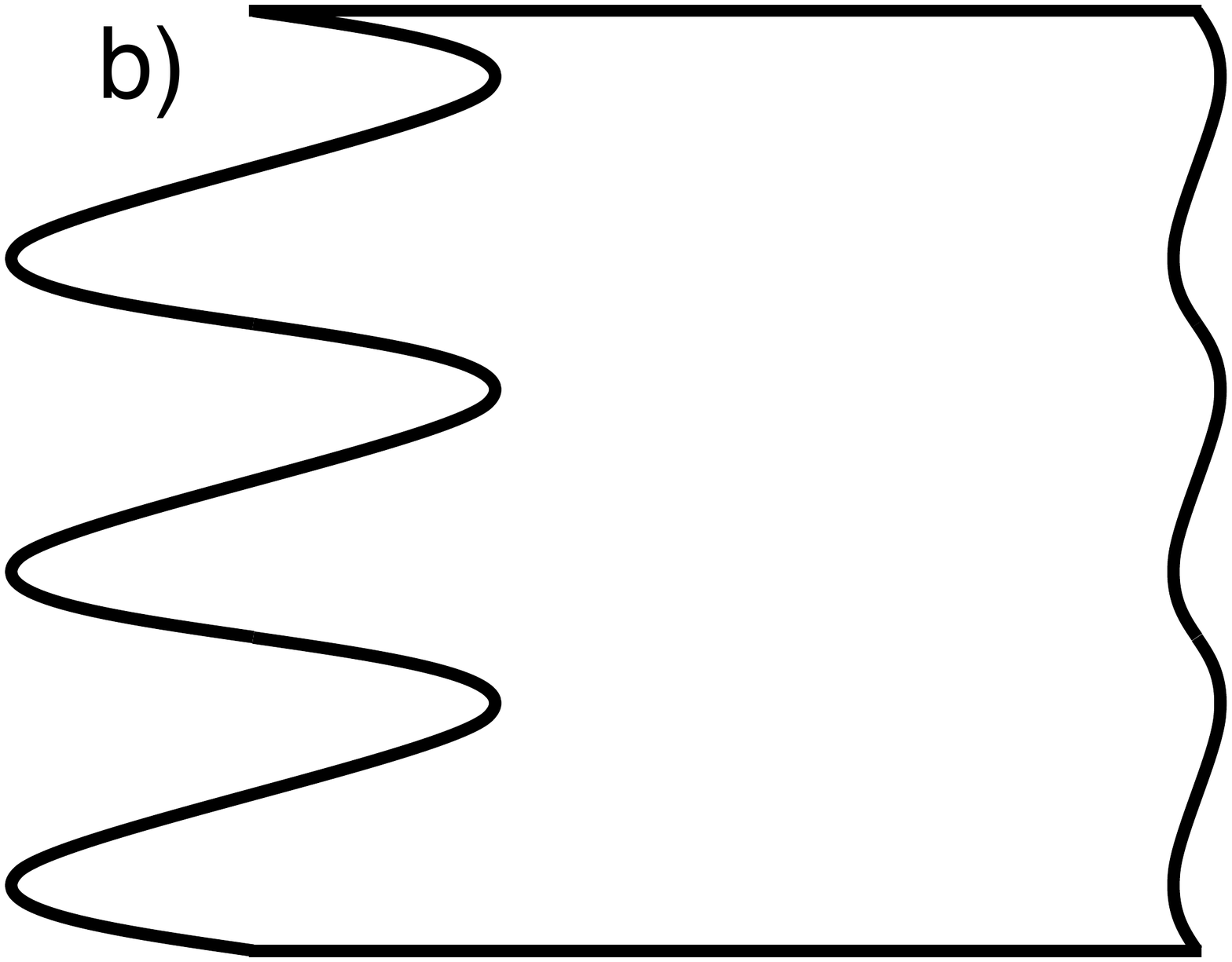}
\includegraphics[width=0.5\textwidth]{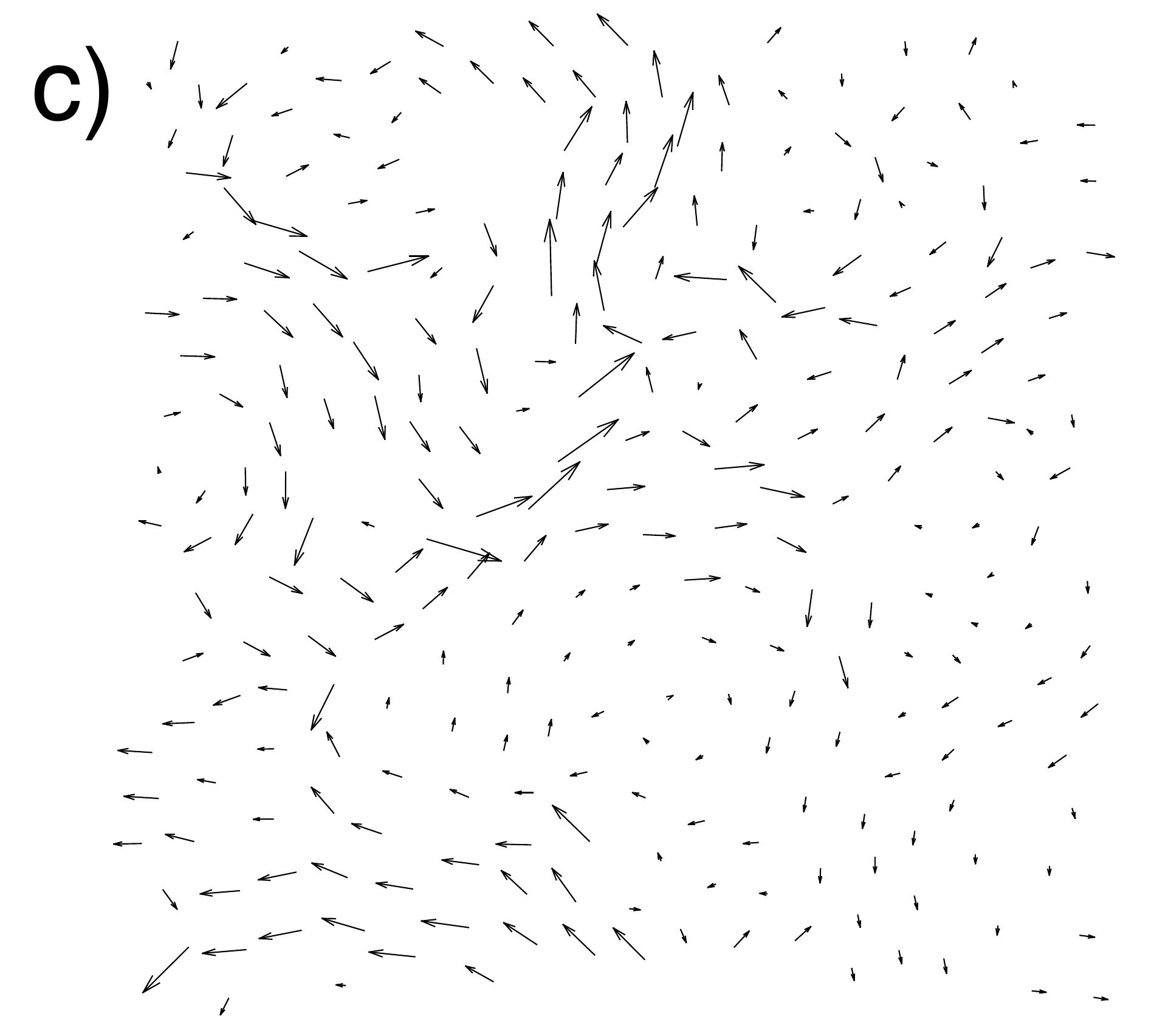}
\includegraphics[width=0.5\textwidth]{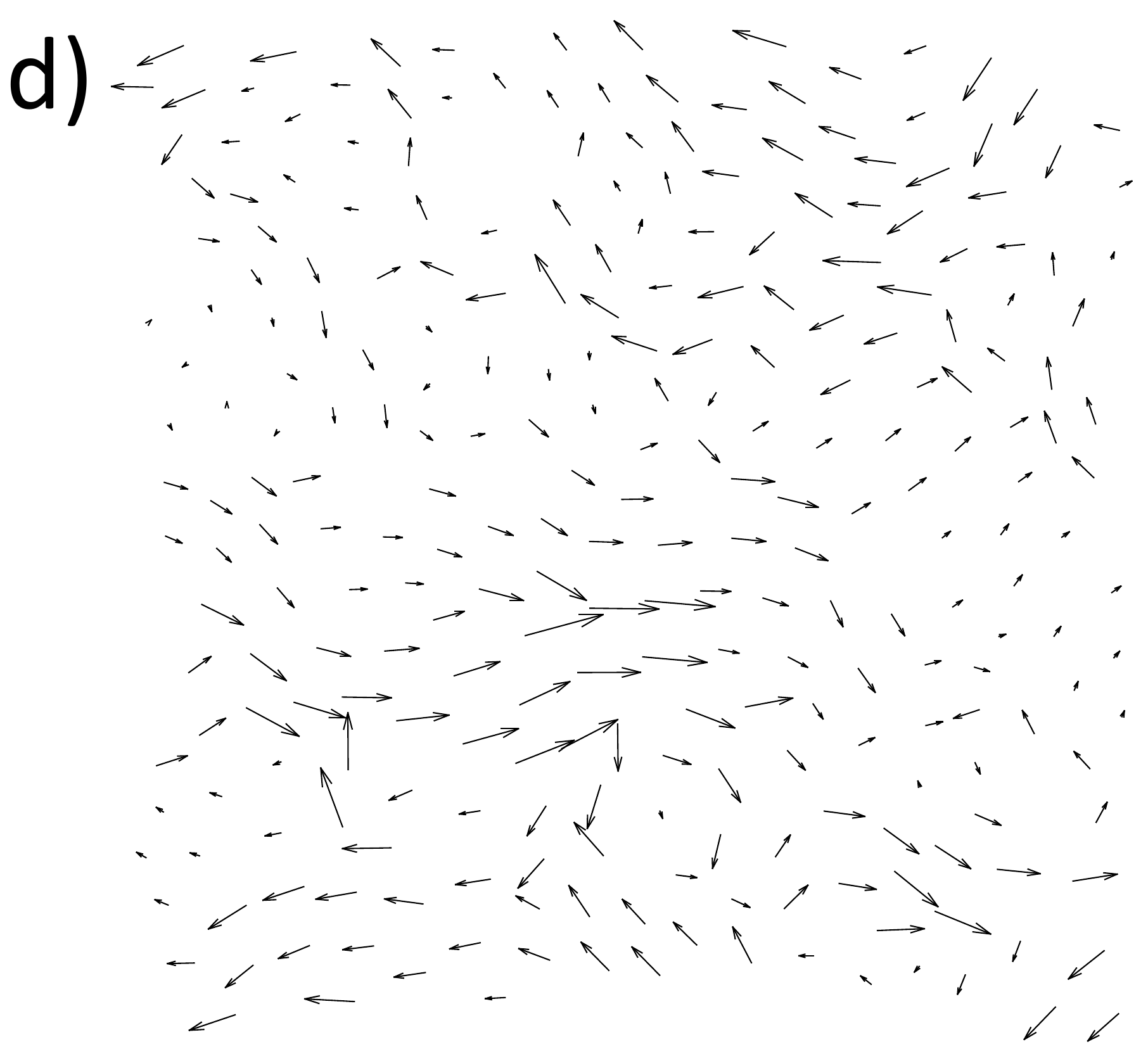}

\caption{Comparison of displacement profiles for elastic and granular bodies.  The sketches in a), and b) indicate decay of sinusoidal displacements imposed at the left side of an isotropic elastic medium\tw{\cite{LandauElasticity}} for two different wavelengths. Panel a) has wavelength equal to the sample width $W$ indicated.   The amplitude at the righthand boundary is reduced as indicated in Eq. \eqref{eq:elasticProfile}.  Panel b) has wavelength  $W/3$.  Amplitude at the righthand boundary is much smaller than in a).  Panels c) and d) illustrate responses to imposed displacements on a marginal granular medium constructed as described in Section \ref{sec:preparation}.  Arrows at left of panel c) show a sinusoidal modulation with wavelength $W$ as in a) .  Displacements in the interior do not retain this sinusoidal form and they are much smaller on the right side as compared to a). Panel d) shows another perturbation with higher wavenumber content and comparable amplitudes on left and right boundaries.  }   \label{fig:transmissibility}
\end{figure}

This leads to a response to forcing that is very different from that of an elastic material. In an elastic material\tw{\cite{LandauElasticity}}, perturbations with long wavelength fall off slowly with distance from the perturbation site, while perturbations with short wavelength fall off quickly.  No such correspondence between wavelength and decay length is present in marginally-jammed granular materials. Figure \ref{fig:transmissibility} demonstrates the difference between these cases. We can then think of transmitting signals through the two materials. A signal can be characterised by a set of scalars, corresponding to amplitudes of each of the modes in a particular basis. We ask, is the dimension of the signal space qualitatively larger in the granular material than in the elastic material?

In fact, we find that it is not. We find an appropriate basis for the space of zero-energy motion in which the system responds to perturbations, and calculate the rate at which these basis modes decay with distance. We show that the decay rates in a marginally-jammed granular material are qualitatively identical to those in an elastic material. This means that signals are transmitted qualitatively no better through a granular material than an elastic material, despite the existence of long-ranged zero-energy motion.

In Section \ref{sec:preparation}, we outline the computational techniques we use to prepare marginally-jammed packings, noting differences in our approach from those generally used. In Section \ref{sec:basis} we discuss how the relevant basis may be formed, while Section \ref{sec:dependence} presents the dependence of mode decay rates on mode number and system width, making comparison in each case with the reference case of an elastic material. In Section \ref{sec:discussion} we present a simple argument for why the observed exponential decay is what should be expected in systems with the geometry we have used. We conclude with a discussion of how this work may be extended, especially to the case of circularly symmetric rather than periodic geometry, to which our simple conceptual argument does not apply.

\section{System preparation} \label{sec:preparation}
\subsection{Jamming procedure}
In constructing our system of particles, we aim to reproduce the well-studied model critically-jammed system of Ref. \cite{OHern2002}. We use $N$ particles indexed by $i$, with radii $a_i$ drawn from a uniform distribution between 0.5 and 1. These particles are placed randomly in a two-dimensional rectangular box of fixed aspect ratio, with side lengths chosen to achieve a packing fraction $\phi$  of 0.82, slightly below the jamming threshold \tw{at which all motions of the contacting particles are constrained}. A marginally-jammed state is then achieved using a two-stage procedure.

First, the particles are taken to have pairwise harmonic repulsion when they are separated by less than the sum of their radii, and no interaction when they are more distant \cite{OHern2002}. The particles' potential energy is minimised by displacing them using gradient descent.  We terminate this minimisation when the maximum displacement step falls below $10^{-5}$. Since this procedure was used as an initialization step, not to find final packings, this value was chosen for convenience based on computational running time.  

This method is efficient at finding a low energy configuration, but not an actual minimum. We depart from previous work \cite{OHern2003} in our technique to find a minimum: our technique exploits linear programming methods, which have been used to test whether or not hard-particle packings are jammed \cite{donev2004linear}. We make use of two vector quantities defined on different spaces. We denote particle displacements from the previously found positions by $\drr$, the vector consisting of all the particle displacements $\{\delta r_i\}$. We denote the separation of particle centers by $\mathbf{l}$, indexing pairs of particles by $\alpha$. Thus any designated pair index $\alpha$ refers to a particular pair of particles,denoted $i(\alpha)$ and $j(\alpha)$ with $j > i$.  Accordingly,  $\mathbf{l}$ has $N(N-1)/2$ entries, though only pairs of particles that are nearby have any physical relevance and are included in the computations. Given the particle positions $\{r_i\}$, we may calculate the compatibility matrix $\CC$ \cite{Kapko2009}, that maps infinitesimal particle displacements to corresponding changes in particle separations by
\begin{equation}
\mathbf{\delta l(\delta\rr}) = \CC~\drr. \label{eq:compatibility}
\end{equation}
We note that $\CC$ is not square. We then seek a set of displacements $\drr$  that eliminates overlaps between particles as determined by the linearized Eq. \eqref{eq:compatibility}, \ie 
\begin{equation}
\delta l_\alpha(\drr) \geq a_{i(\alpha)}+a_{j(\alpha)}-l_\alpha \label{eq:ineq}
\end{equation}
for all pairs $\alpha$.
In order to satisfy Eq. \eqref{eq:ineq}, pairs that are separated by less than the sum of their radii must move apart by the shortfall; pairs that are separated by more than the sum of their radii are permitted to move closer together, by up to the size of the remaining gap between them.  A standard linear programming solver (Matlab's \linprog\cite{Matlab}) is used to find a solution $\drr^*$ to this set of inequalities, or determine that it does not have a solution.   If a solution exists, it is part of an allowed region, and this region defines where particles may be placed without overlapping (under the approximation of linearity).   Within this region, we seek a displacement $\drr'$ with smaller magnitude than $\drr^*$ when this magnitude is nonzero.  To this end we define $S$ to be the set of $\alpha$'s for which the inequalities in \eqref{eq:ineq} are in fact equalities.  That is, 
\begin{eqnarray}
\delta l_\alpha(\drr^*) &= a_{i(\alpha)}+a_{j(\alpha)}-l_\alpha\quad , &\alpha\in S \label{eq:justexact} \\
\delta l_\alpha(\drr^*) &> a_{i(\alpha)}+a_{j(\alpha)}-l_\alpha\quad , &\alpha\notin S. \label{eq:juststrict}
\end{eqnarray}
If $\drr^*$ were an arbitrary point satisfying \eqref{eq:ineq}, we would expect $S$ to be empty, but since $\drr^*$ was found by a linear programming algorithm, in fact many inequalities are equalities. The $\alpha$'s in $S$ represent pairs of particles in contact.

Now, the solutions of \eqref{eq:justexact} define the intersection of hyperplanes in the $\drr$ space which includes $\drr^*$.  We may readily find other $\drr$'s that satisfy \eqref{eq:justexact} but are smaller in magnitude than $\drr^*$.  We denote the smallest of these $\drr$'s as $\drr_1$.  It is simply the projection of $\drr^*$ perpendicular to the  intersection of hyperplanes representing solutions to \eqref{eq:justexact}. In general, a subset of the strict inequalities of Eq. \eqref{eq:juststrict} are also satisfied by $\drr_1$.  However, some of the conditions in \eqref{eq:juststrict} are violated by $\drr_1$.  To remove these violations we then repeat the \linprog procedure under further restrictions: all the contacts in $S$ must remain in contact; in addition, all the overlapping pairs in $\drr_1$ must become contacts, satisfying \eqref{eq:justexact}.  The full set of contact conditions is denoted $S_1$.  The result is some configuration denoted $\drr^*_1$ in which all pairs of particles that were in contact in $\drr^*$ remain in contact, all pairs of particles that overlapped at $\drr_1$ are also in contact, and all other pairs of particles are not overlapping (they may be in contact or separated). As in the initial iteration, the procedure usually creates additional contacts (that is, pairs of particles for which the constraint is an equality instead of a strict inequality). Then as before, we determine the smallest $\drr$ that retains all existing contacts.  This we denote $\drr_2$. Again the passage from $\drr^*_1$ to $\drr_2$ creates overlaps for some additional pairs.  We again remove these overlaps by rerunning \linprog, requiring preservation of existing contacts and conversion of all overlaps in $\drr_2$ to contacts.  Each such iteration increases or maintains  the number of contacts.  But this number cannot increase indefinitely.  At some iteration $i$ the \linprog step fails or else the passage from $\drr_i^*$ to $\drr_{i+1}$ produces no overlaps.  Such a $\drr_{i+1}$ satisfies the initial non-overlap conditions \eqref{eq:ineq} and also has a magnitude that is locally minimal.  This procedure gives a candidate for a minimal non-overlapping state, denoted $\drr^*\mmin$.  

Once $\drr^*\mmin$ has been determined, the particle positions $\rr$ are displaced by this $\drr^*\mmin$ and new $l_\alpha$'s are computed. The new state may have overlaps, since $\drr^*\mmin$ only removed them in a linearized approximation.  Thus the process is repeated until there are no overlaps  (to a tolerance of $10^{-13}$ per particle in energy). 

Such non-overlapping configurations can readily \tw{be} found at low packing fractions $\phi$.  These are not in general jammed.  To attain a jammed state we now compress the system uniformly so as to increase the packing fraction.  Thus all contacts become overlaps. Then the above procedure for finding a configuration without overlap is repeated. When instead the linearised inequalities are found to have no solution (i.e. \linprog fails), the current step is discarded, the compression step size decreased and a new attempt taken from the previous non-overlapping configuration. This procedure is repeated until the step size falls below a relative compression of $10^{-9}$. This configuration is our representation of a marginally jammed state.  We found the linear programming method to be much faster and more reliable than a simple implementation of energy minimisation\footnote{In contrast to typical linear-programming problems, we do not require minimisation of any ``objective function" in our method.}.

\begin{figure}
\begin{center}
\includegraphics[width=.8\textwidth]{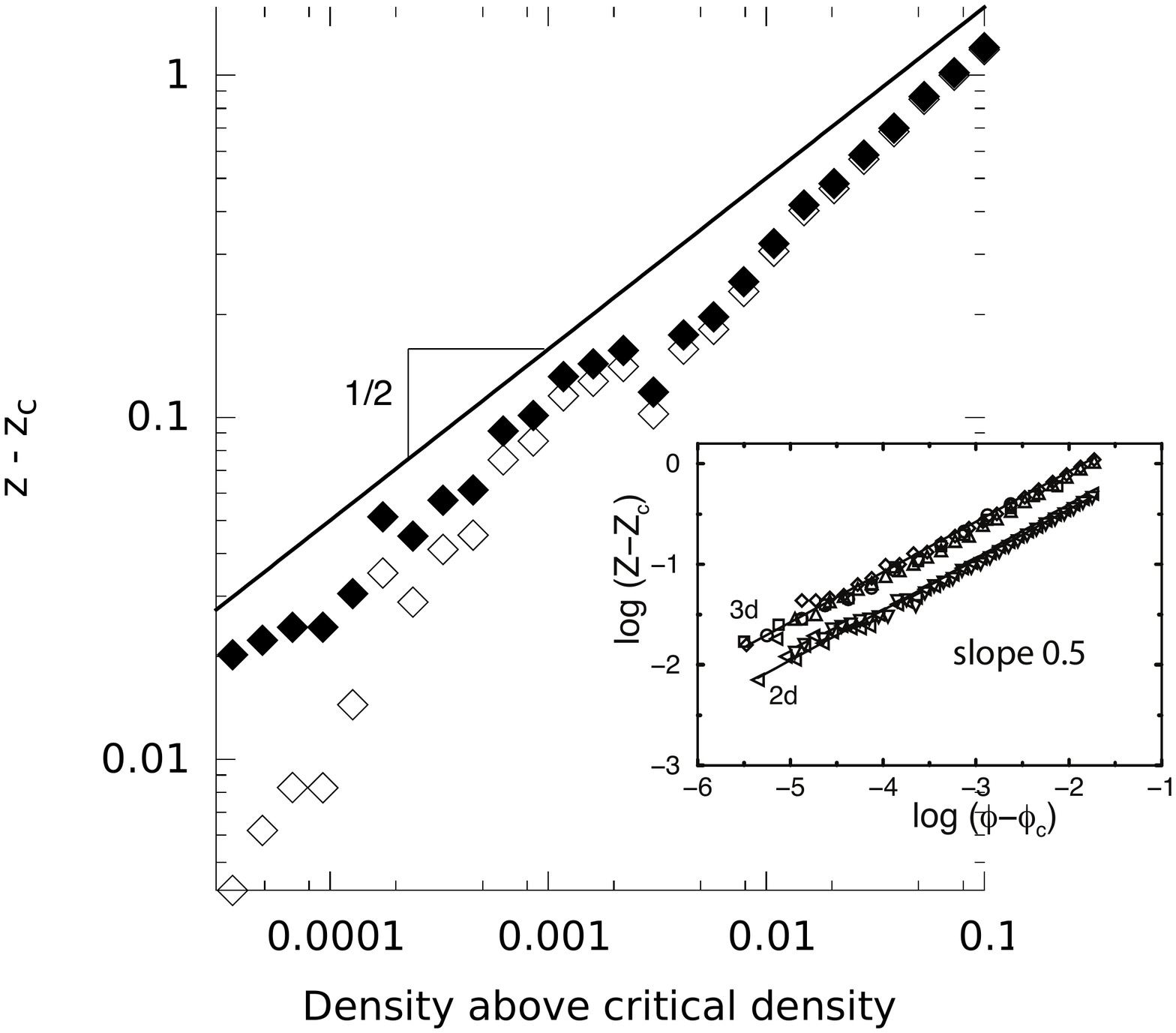}
\includegraphics[width=0.5\textwidth]{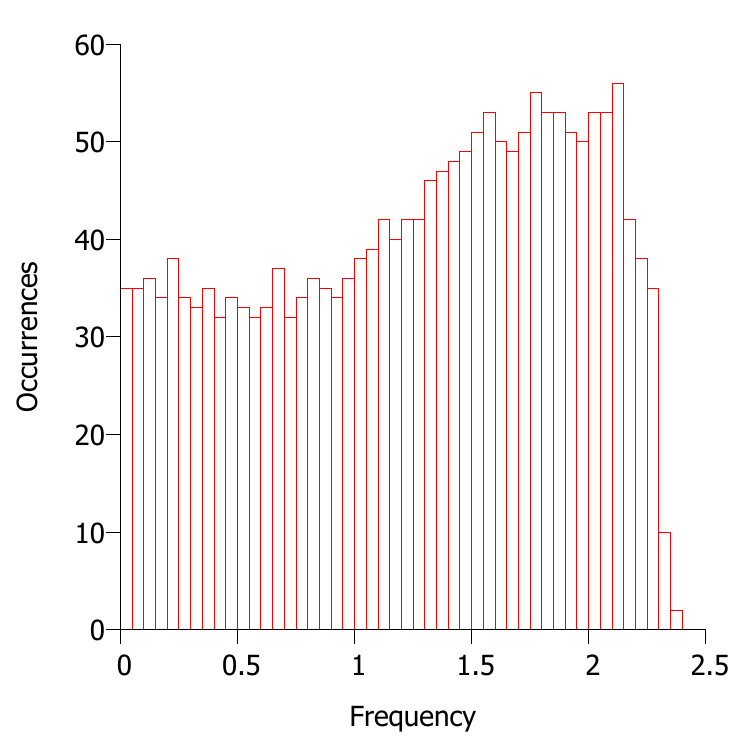}\vbox{\hbox{\includegraphics[width=.35\textwidth]{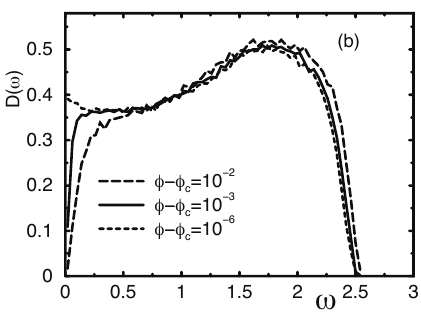}}\vskip .3in}
\caption{Top: Excess co-ordination number $z - z_c$ as a function of density above the critical jamming density.  Inset is from Fig. 9 of Ref \cite{OHern2003}, where $z_c = 3.98$ was chosen for best fit to a power-law form.  Main graph is from present work using a  system of 1024 particles. Solid symbols: $z - z_c$ using the same value of $z_c$ as in inset.  Open symbols: excess above the isostatic number $z_c = 2(2M - 2)/M$ discussed in the text.  Bottom: Histogram depicting the density of states of the normal modes of a marginally-jammed system of 1024 particles from the present work. Inset: analogous plot from Fig. 12b of Ref. \cite{OHern2003}}\label{fig:cfOhern} 
\end{center}
\end{figure}

We verified that our system resembles the jammed systems used to establish scaling properties of jamming\cite{OHern2003}.  To this end we further compressed the system, minimising energy by gradient descent and noting the additional number of contacts per particle as a function of density above the critical jamming density. Results, plotted in Figure \ref{fig:cfOhern}, are similar to those of Ref. \cite{OHern2003}; exact agreement is not necessarily expected as our particle radii are drawn from a different distribution.  For further verification we computed the density of vibrational mode frequencies defined in \cite{OHern2003}.  We also determined the inverse participation ratios  of the single-bond modes defined in \cite{vanHecke2010}, Sec 3.5.1.  Comparisons are shown in Figure \ref{fig:cfOhern}.  In all these tests our system shows the distinctive features of the marginally-jammed systems in the literature.

After discarding rattlers (particles with fewer than three contacting neighbours), leaving $M$ particles in the jammed cluster, we form the compatibility matrix of the final zero-energy state, considering only  pairs $\alpha$ that are touching (i.e. separated by exactly their sum of radii, to within a small numerical tolerance). We then check whether the number of touching pairs is $2M-2$, as required by Maxwell counting \cite{Maxwell1864}: if it is not, we discard the packing as not truly at the jamming transition.  \tw{This discarding can happen as much as 50 percent of the time for the largest packs.}  (it is possible for a critically-jammed configuration to have more than $2M-2$ touching pairs, but this is not generic). To remove the trivial translational degrees of freedom, we add an additional row to $\CC$ with entry 1 in every column that corresponds to the $x$ displacement of a particle. This gives a component of $\mathbf{l}$ that is the average movement in the $x$ direction, so that any set of displacements for which $\mathbf{l} = 0$ (or at least this component of $\mathbf{l}$ is zero) must have zero center of mass motion in the $x$ direction. We add also a similar row for the $y$ direction.
The compatibility matrix $\CC$ is then a square matrix. Its singular values are proportional to the normal mode frequencies: their distribution, plotted in Figure \ref{fig:cfOhern} is similar to those of O'Hern et al. \cite{OHern2003}.

As is commonly found in marginally jammed systems, $\CC$ proves to be invertible. If $\CC$ were not invertible, there would be a set of displacements $\drr^0$ that did not change any particle separations, i.e. a zero-energy mode of motion. We generically expect that movement in either the positive or the negative direction of this motion would allow the box size of the packing to be decreased while maintaining zero energy, and so by definition the system preparation would not be complete.

\begin{figure}
\begin{center}
\includegraphics[width=0.5\textwidth]{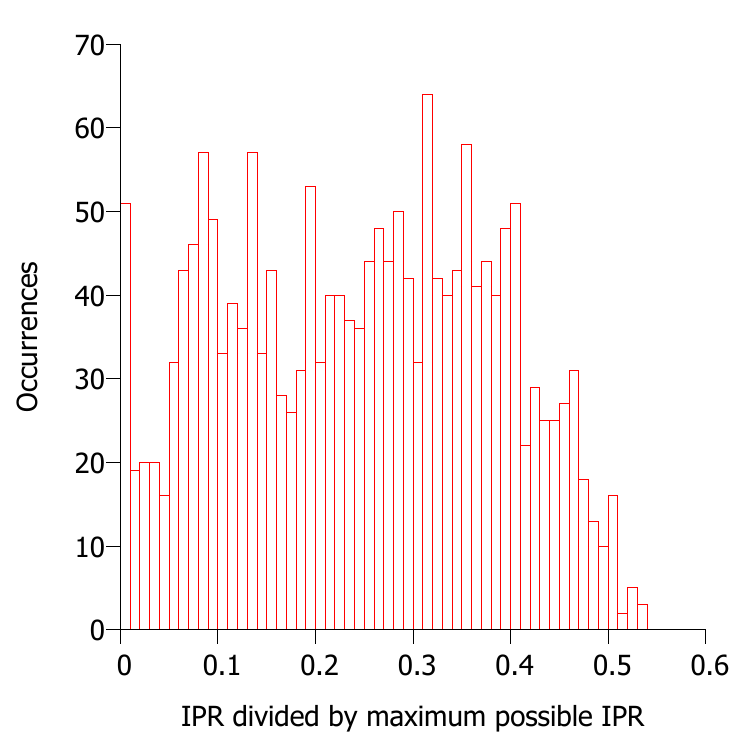}
\caption{Histogram of inverse participation ratios (\cite{vanHecke2010} Sec 3.5.1) for free modes created by cutting each single bond from the packing used for Figure \ref{fig:cfOhern}.}
\label{fig:invparticipation}
\end{center}
\end{figure}

Each column $\alpha$ of $\CC^{-1}$ gives the unique displacement mode $\drr$ corresponding to a slight increase in the separation of the pair $\alpha$ while leaving all others unchanged. Considering the network of particles and touching pairs to be a network of nodes and springs, it is the free linear mode of motion that becomes available if bond $\alpha$ is deleted. The spatial extent of these modes is proportional to the size of the system \cite{Wyart2005,Lerner2013,Muller2015}, as can be seen in the histogram of inverse participation ratios shown in Figure \ref{fig:invparticipation} for $N = 1024$.

\begin{figure}
\includegraphics[width=0.71\textwidth]{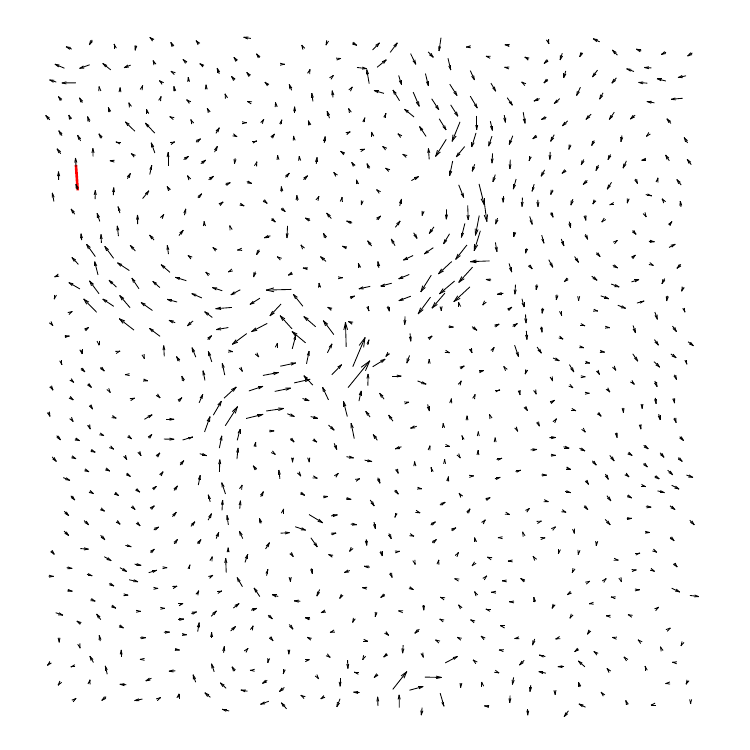}
\includegraphics[width=0.71\textwidth]{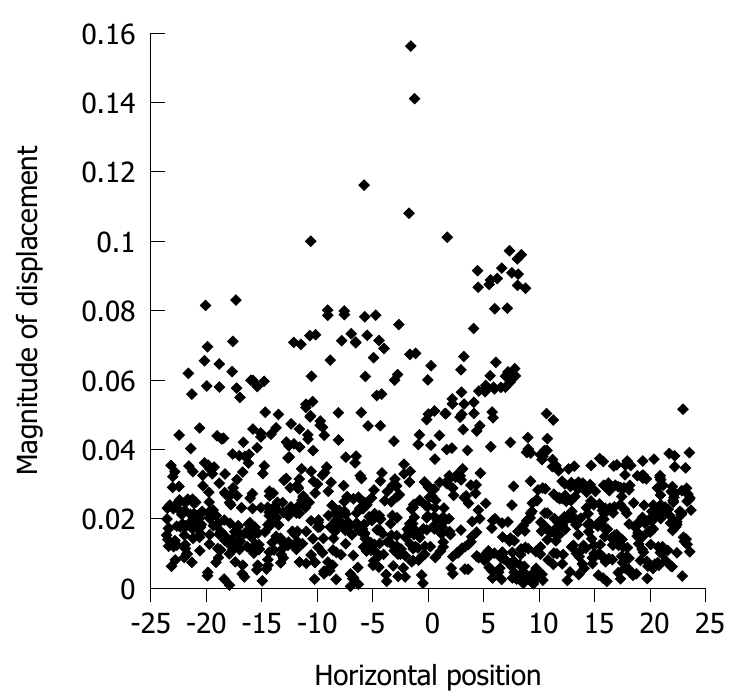}
\caption{(a) Relative displacements of particles in the zero-energy mode created by removing one constraint, shown as a colored bond line, from a marginally-jammed system of 1024 particles. (b) A projection of the same information plotted in (a): the magnitude of displacements, as a function of position in the horizontal direction only.  \label{fig:quiver}}
\end{figure}

Figure \ref{fig:quiver} shows a typical mode coming from removing one such constraint, represented in two ways: the vector displacement of each particle as a function of its position in two dimensions, and the magnitude of displacement as a function of position in the horizontal direction only.

\subsection{Imposition of boundary conditions}
Our goal is to determine whether a perturbation in one region of an isostatic system can be characterised by making measurements in a distant region. To test this, we remove the periodicity in the horizontal direction by deleting all bonds that cross a particular vertical line.  The number $m$ of deleted bonds is proportional to the width of the system.  This results in an underconstrained system: $\CC$ has fewer rows (constraints) than columns (degrees of freedom). Thus the system has numerous linearly independent modes of motion allowed at zero energy. An assessment of the degree of localisation of these modes is needed to determine whether perturbations can be detected at long distances. 

\section{Mode localisation}
\subsection{Basis of decaying modes} \label{sec:basis}
Our overall aim is to assess the degree to which different perturbations applied at one side of the granular material can be distinguished at the other. As shown in Figure \ref{fig:cfOhern}, cutting a single bond typically produces a mode that spans the entire material, suggesting that any imposed perturbation could at least be detected anywhere in the material. But it is a stronger statement to say that different modes can be distinguished: this requires that the long-ranged parts of the different modes are not ultimately the same. To determine whether this is the case, we need to form a suitable orthogonal basis.

In identifying the appropriate basis in which to assess possible localisation of the effect of perturbations, we can take inspiration from an isotropic, linear elastic material\tw{\cite{LandauElasticity}} with the same shape as our system: periodic in the vertical direction, with free boundaries in the other direction as shown in Figure \ref{fig:transmissibility}a, b. For simplicity, we consider the case where the distance between the free boundaries is of the same order as the width. The perturbations that can be put into one side are continuous and hence are spanned by an infinite-dimensional basis. We could choose from many different bases, but a natural choice is one of sinusoidal waves. These are periodic in the vertical direction with wavenumbers \tw{}$k = 2\pi n/W$, where $W$ is the width  of the two free faces and \tw{}$n$ is an integer: Figure \ref{fig:transmissibility} illustrates some perturbations from this basis \tw{and defines co-ordinates $x$ and $z$}. When such a perturbation is applied at one side, the displacement at long distances in the interior of the material decays exponentially \tw{for large} $z$:
\begin{eqnarray}
\label{eq:elasticProfile}
\delta \vector r(x, z) = e^{-\kappa z}~
\left[ (\vector A  \cos kx + \vector B \sin kx ~) + \kappa z~(\vector B' \cos kx + \vector A' \sin kx~) \right ] \quad\quad
\end{eqnarray} 
with decay rate $\kappa$ equal to $|k|$.  The \tw{vector} coefficients $\vector A$ and $\vector B$ are determined by the imposed surface displacement.  The coefficients $\vector A'$ and $\vector B'$ are linear combinations of $\vector A$ and $\vector B$ determined by the Poisson ratio of the material. (See e. g. \cite{Sadd:2014fk,LandauElasticity} \tw{Thus there are four independent modes (of amplitudes $A_x$, $A_z$ and $B_x$, and $B_z$) for each distinct $\kappa$, and to accommodate $p$ modes requires on average $n = p/4$ or $\kappa = (p/4)~(2\pi/W)$.}
Now, if measurements can be made with precision $\epsilon$ relative to a unit input, a particular mode can be detected at distance $R$ into the material only if $e^{-|k|R} > \epsilon$. The number of modes \tw{}$p$ that can be detected at this distance, and hence the dimension of the space in which signals could be sent using these perturbations, is thus \tw{}
\begin{equation}
p = -\frac{2W}{\pi R}\ln\epsilon.
\end{equation}

It is important to note that, although this calculation was performed in a particular basis, a greater number of distinguishable modes cannot be found simply by working in a different basis. For example, if we were to take a detectable mode $\psi_1$ and an undetectable mode $\psi_2$, and form new modes $\psi_A = (\psi_1+\psi_2)/\sqrt{2}$ and $\psi_B = (\psi_1-\psi_2)/\sqrt{2}$, both $\psi_A$ and $\psi_B$ would have sufficiently large magnitude at $R$ to be detected. However, they would not be distinguishable from each other: both would appear to be equal to $\psi_1/\sqrt{2}$. Thus these modes would still contribute only one to the number of distinguishable modes. In general, one must work in what may be termed the most ``pessimistic'' basis: the one in which modes decay as rapidly as possible. 

In our system constructed from a granular material, we wish to characterize the $m$ free modes created by removing boundary bonds.  We do not expect the useful basis modes to be as simple as sinusoids.   
Nevertheless, we may construct basis modes maximally concentrated on the left or right side.
In order to distinguish these modes clearly, we again take a system whose depth is comparable to its width. We first determine the normalized mode with the smallest intensity on the right hand side denoted mode 1L.  For definiteness, we define this intensity as the sum of squared amplitudes of the $2m$ particles farthest to the right
\footnote{The exact number $2m$ is not very important: it was chosen because there are approximately $m$ particles in what might be termed one layer, and the outermost layer sometimes has slightly unusual behaviour due to the asymmetry of bonds to it, so using $2m$ particles ensures that the next layer is included.
         }. 
Next we find  the mode (denoted 1R) orthogonal to mode 1L having the smallest intensity on the {\em left} side.  For subsequent modes we choose the side where the last mode constructed had smaller intensity and build the mode orthogonal to the existing modes with lowest intensity on that same side.  The result of this process is an orthogonal basis consisting of left-concentrated modes 1L, 2L... and right-concentrated modes 1R, 2R, ... in roughly equal numbers.  We then label these modes by a label $p$ analogous to the sinusoidal modes of an elastic material. The final (least-concentrated) mode in the basis is designated $p=0$.  The left-concentrated modes are then assigned successive labels with $p = 1, 2, ... p\mmax$ following the inverse order in their construction.  Likewise the right-concentrated modes are assigned negative $p$ values from $-1$ to $p\mmin$.   Thus the sequence from $p\mmin$ to $p\mmax$ proceeds from the mode most strongly concentrated on the right to the modes most strongly concentrated on the left.

\begin{figure}
\includegraphics[width=0.5\textwidth]{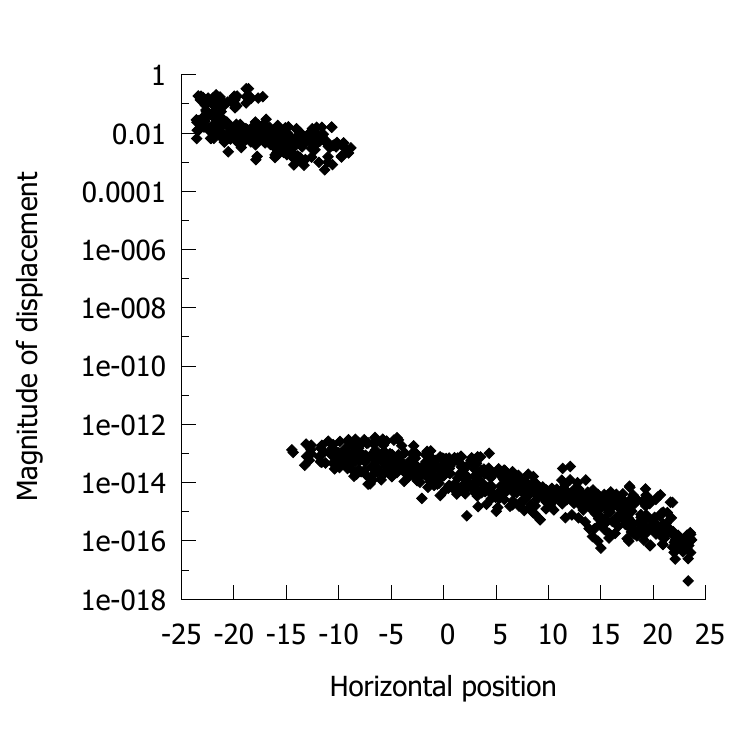}
\includegraphics[width=0.5\textwidth]{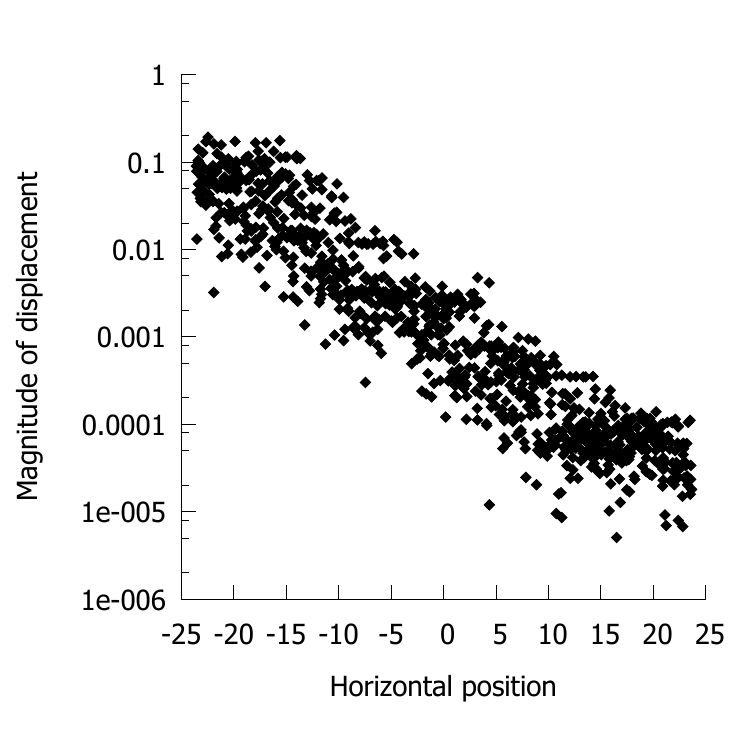}
\includegraphics[width=0.8\textwidth]{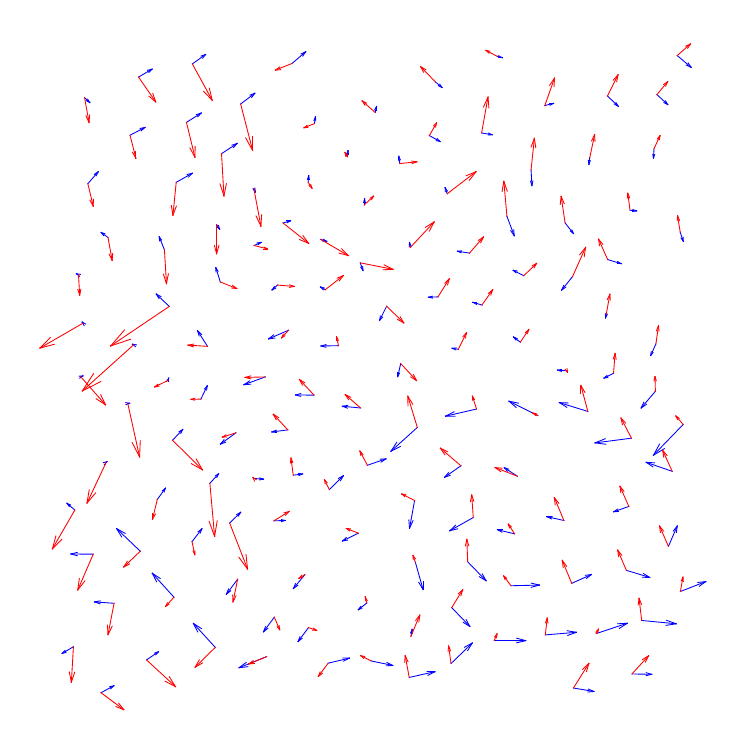}
\caption{(a) Magnitude of displacement vs horizontal position for a basis mode that is zero over a region at large $x$, in a system of 1024 particles. (b) An equivalent plot for basis mode that decays exponentially with $x$ without becoming exactly zero. (c) Vector displacements of particles in each of the two modes that decay most slowly, for a system of 144 particles. \label{fig:basismodes}}

\end{figure}

Two kinds of basis state result from this procedure. First, some states are non-zero only on a restricted domain, and the displacement of particles outside this domain is numerically zero. This indicates that the bond-cutting procedure creates regions that are underconstrained, even if the rest of the system is considered to be rigid. These finite modes can certainly not be used to gain information by making measurements at long distances. The second type of state is non-zero over the entire sample, but the typical displacement of a particle depends exponentially on position in the horizontal direction. This shows that the transmission of some perturbations resembles that in an elastic material. Examples of the two types of mode are shown in Figure \ref{fig:basismodes}. We show also an example of the two slowest-decaying modes. There is no particular correlation between the displacement of particles in one of the slowly-decaying modes and its displacement in the other. Each mode is heterogeneous and does not resemble a single Fourier mode. 

\subsection{Dependence of decay rates on mode number and system width} \label{sec:dependence}
\begin{figure}
\includegraphics[width=\textwidth]{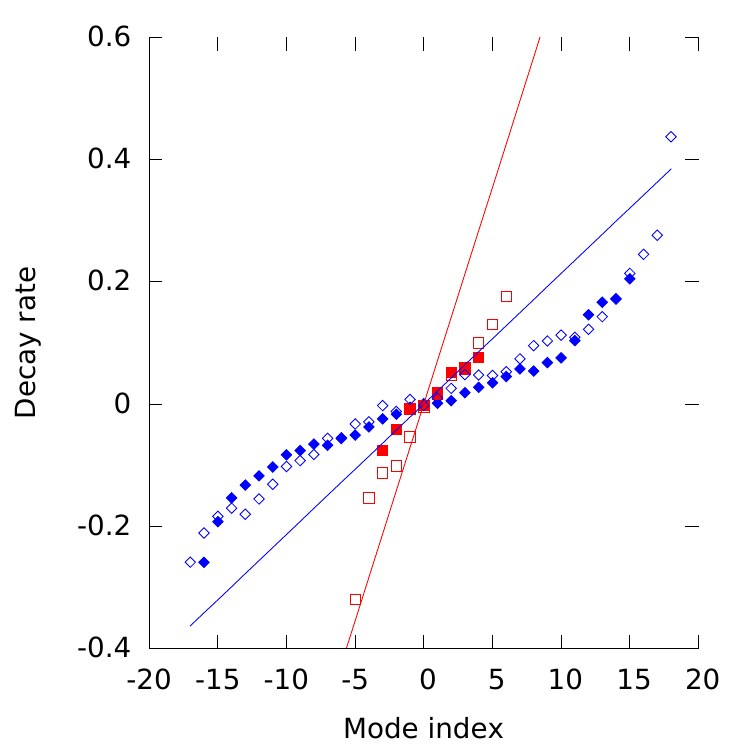}
\caption{Decay rates $\kappa$ as a function of mode index for systems with widths \tw{$W$ of 22.2} (red squares) and \tw{73.5} (blue diamonds) and depths of \tw{29.5} (open symbols) and \tw{59} (filled symbols), along with the linear dependence of decay rate with mode index for elastic materials of the same widths (lines), \tw{from Sec. \ref{sec:dependence}.  Length unit is the radius of the largest \tw{disks} in the pack.} \label{fig:decay}}

\end{figure}

We perform a least squares fit to find the decay constant $\kappa$ of each exponentially decaying mode, where the unit of length is the maximum particle radius. Since there are modes located on each side of the system, the calculated decay constant can be positive or negative.  Figure \ref{fig:decay} shows that decay constants of the exponential modes depend linearly on $p$, with a slope that depends on system width (depicted in the vertical direction throughout this paper) but not depth (horizontal). This behaviour is the same as that of elastic materials.  For comparison we show the dependence of $\kappa$ on mode number \tw{$p$} for elastic media, \tw{as explained below Eq. \ref{eq:elasticProfile}---\viz  $\kappa = (\pi/(2W))~p$.}

\begin{figure}
\includegraphics[width=\textwidth]{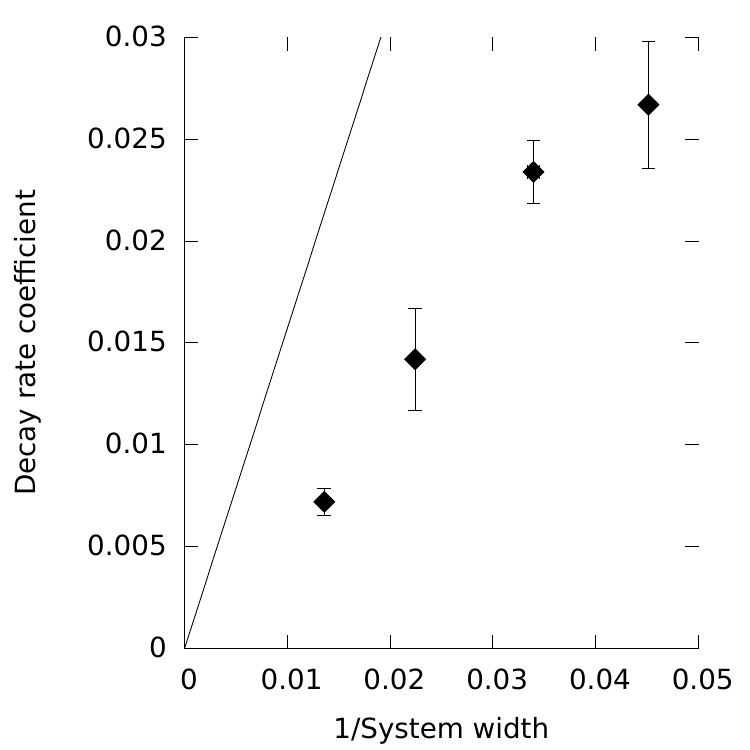}
\caption{Increase in decay rate per mode from slopes near $p=0$ taken from plots like Figure \ref{fig:decay}. vs reciprocal system width $1/W$. Length unit is the radius of the largest \tw{disks} in the pack.   Straight line indicates the behavior of an isotropic elastic medium, like the straight lines in Figure \ref{fig:decay}.
}   \label{fig:widtheffect}
\end{figure}

Since the decay rate depends linearly on mode index $p$, we can calculate a decay rate coefficient $\chi$ by fitting observed decay rates to the function
\begin{equation}\label{eq:chi}
\kappa = \chi p+\kappa_0.
\end{equation}
Figure \ref{fig:widtheffect} plots $\chi$ as a function of the reciprocal of the width of the system, averaged over different system depths and different realisations at the same depth. These granular systems show a linear dependence of decay rate coefficient on the reciprocal of system width: $\chi \approx 0.6/W$. This means that the $p=1$ mode will decay exponentially with characteristic length $W/0.6$, the $p=2$ with length $W/1.2$, and so forth. If, for example, a sender wished to send a signal comprising five scalars to a depth equal to the width of the system, the receiver would need to be able to measure with a precision of $e^{-.6 \times 4} = 0.09$ (remembering that the $p=0$ mode is available). For low-dimension signals, detection does not require unusual precision, due to the small value of the constant 0.6. The important point is that the required detection precision increases exponentially with the depth of the system or the dimensionality of the signal, and does not decrease if the system is made wider with fixed aspect ratio.

The modes decay more slowly than in an elastic \tw{material\tw{\cite{LandauElasticity}}}, where \tw{$\chi = \pi/(2W)$}. Many factors could contribute to this difference. One that we believe is particularly important is the uncertainty in the definition of the most suitable basis modes. In the limit of an extremely deep system calculated with perfect precision, the fastest-decaying mode is unambigious: it is that mode that has the highest exponential decay constant in the limit of large depth. After setting this mode aside, the fastest-decaying mode in the perpendicular subspace can be identified in the same way. Modes could be just as easily identified from the slow-decaying end. In a finite system, however, it is not possible to perfectly identify the fastest-decaying mode. The method used tends to overestimate its decay rate, hence underestimating the decay rates of the slowly decaying modes that are used in calculating $\chi$. This may account for some, though probably not all, of the observed difference between these granular systems and the universal result for isotropic elastic materials.

\section{Discussion} \label{sec:discussion}

By this measure, granular systems at the jamming transition behave in the same way as elastic systems, which are not close to marginal stability and do not have long-ranged modes other than bulk translation and rotation. It is useful to ask what causes this qualitatively identical behaviour, and whether a modified system may be able to display qualitatively different decay of the basis modes.

A way to understand the exponential decay is to imagine cutting the system as constructed into strips, with cuts made perpendicular to the direction of propagation. If this is done, the number of zero-energy modes of motion available to each strip is approximately equal to the number of zero-energy modes of the system as a whole, because this number depends only on the boundary and the boundary conditions are the same. 

Having identified the number of zero-energy modes available in a strip (ignoring those of finite extent), the state of the strip can be defined by fixing the same number of particle displacements at the left hand (sending) boundary: the values of these displacements may be formed into the vector $\mathbf{x}$. The state can also be measured by measuring this number of displacements at the right hand (receiving) boundary: these measurements may be formed into the vector $\mathbf{y}$. The strip can then be described by a transmission matrix $T$ where $\mathbf{y} = T\mathbf{x}$. Since the particles on the right hand side of this strip can be used as the left hand side of an overlapping strip, the entire system will have a transmission matrix that is a product of the transmission matrices of many strips. Exponential decay is thus the natural behaviour to expect of any quantity that depends on depth.  Indeed, it is difficult to imagine a non-exponential falloff without some kind of correlations between $T$'s at different depths.  Thus the sequence of exponentially falling modes that we observed in our system is natural.  

Our findings show an additional regularity beyond simple exponential falloff.  The decay length is found to be linear in mode number, with a coefficient that resembles a uniform elastic material, as noted above.  The reason for this resemblance is not obvious, inasmuch as the modes in our packing are quite nonuniform.  \tw{ This resemblance to elastic materials, found in our marginally-jammed simulations, may be applicable more broadly to other isostatic networks\cite{Mao:2010fk, Lieleg:2011uq} as well.  It would be of interest to investigate such systems.  This property is certainly applicable to frictional granular packings and other overjammed, hyperstatic packings, since these are known to behave on large scales like conventional elastic materials.}

This resemblance between our granular system and an elastic medium suggests that the two systems are mechanically equivalent at large scales.  However, a conventional elastic description dictates a specific  dependence of decay rate on mode number, and our study obtained a contrasting dependence, as shown in Figures \ref{fig:decay} and \ref{fig:widtheffect}.  How should this puzzling contrast be interpreted?  Does it indicate a qualitative distinction between the granular material and an elastic material?  Above, we have suggested a more mundane interpretation.  We noted that numerical difficulties in precisely determining the most strongly localized modes may account for the puzzling contrast in decay rates.  In future work it would be useful to understand this apparent contrast with \tw{an} elastic material. 

There is much recent interest in the localization of modes in isostatic networks, such as the jammed packings of our study. In ordered near-marginal systems \cite{Kane2014}, they typically produce modes strongly localized at a boundary, and hence much more localized than the modes discussed above.  Anisotropic systems such as those prepared under gravity or by perturbing a square lattice, also show behaviour that is significantly different from the systems in this work \cite{Cates1999}, without escaping the regime of exponential decay.

An important difference between granular and isotropic elastic materials is that, in the former, modes do not resemble simple Fourier modes and vary greatly from realisation to realisation.

Throughout this work, we have focused primarily on the \tw{}pessimistic basis, in which modes decay as rapidly as possible from one side of the system to the other. This is the most useful basis for quantifying the distinguishability of perturbations. The modes in any other basis can be thought of as superpositions of pessimistic basis modes, so each mode in the other bases will have a combination of different decay lengths. One could find characteristic decay lengths for these modes: such decay lengths would presumably be distributed over a similar range to those in the pessimistic basis, but weighted towards smaller lengths, as these dominate at large depths. This is consistent with recent observations \cite{Sussman:2016kq}.

\begin{figure}
\includegraphics[width=0.5\textwidth]{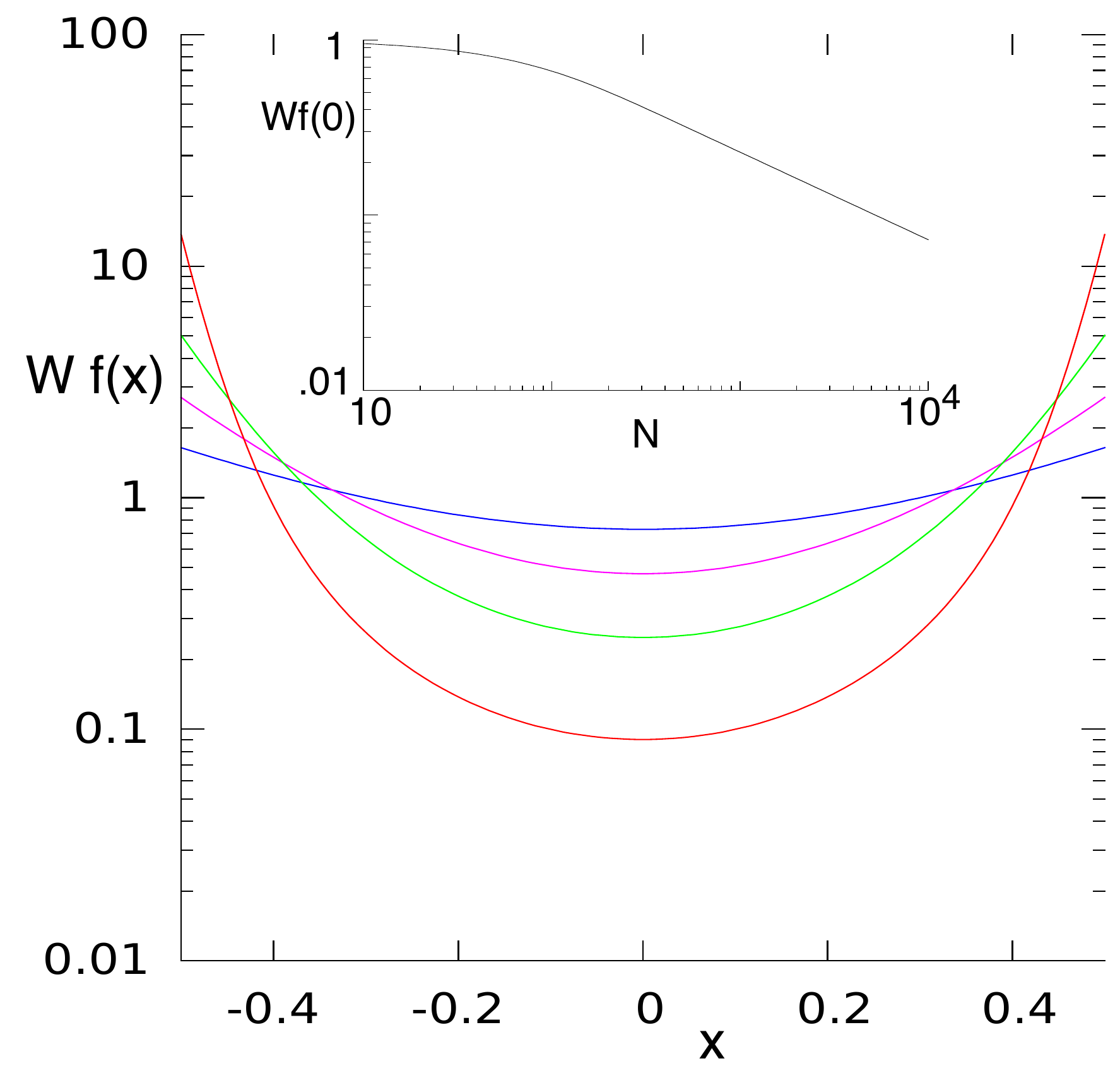}
\vbox{\hbox{\includegraphics[width=.5\textwidth]{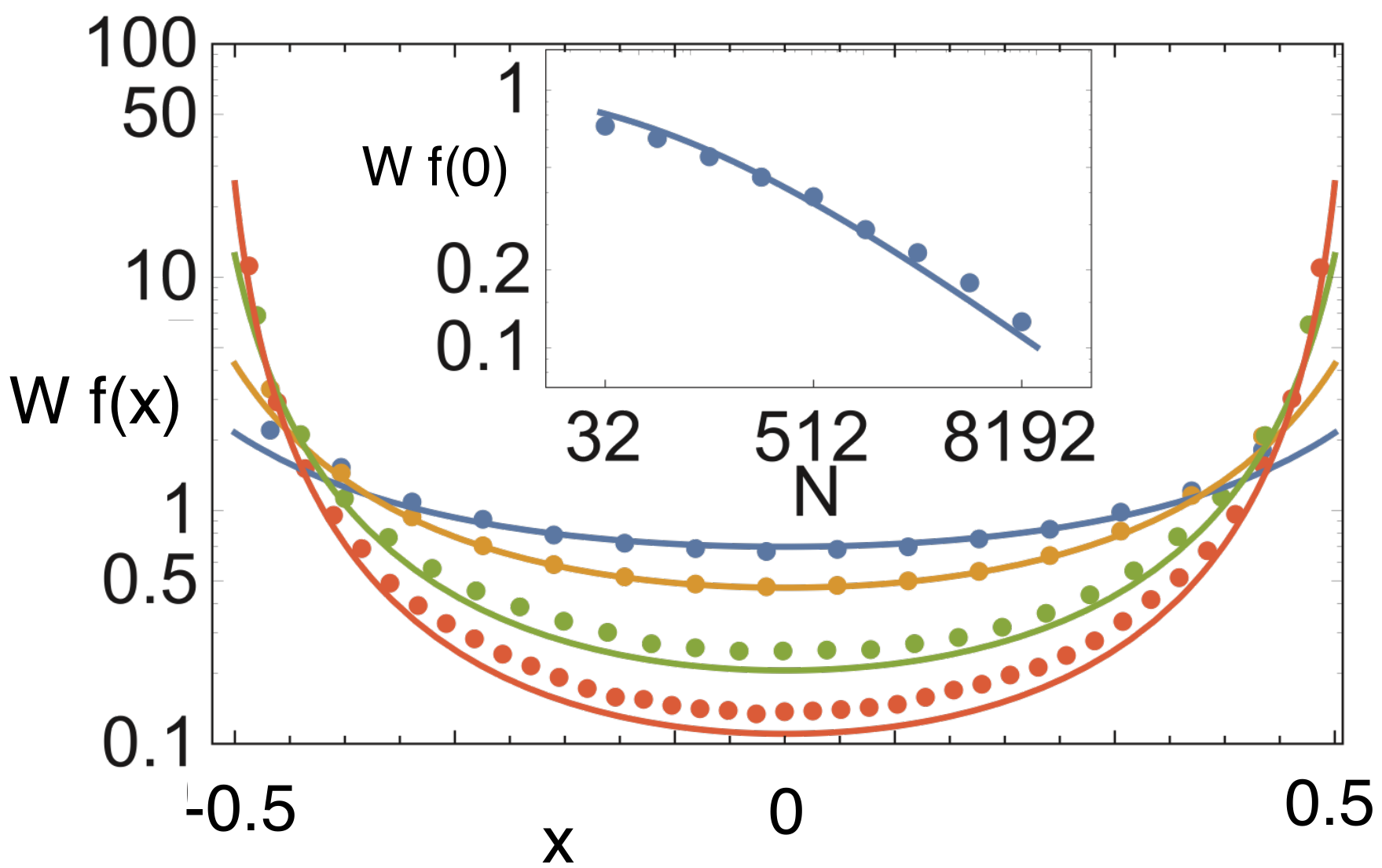}}\vskip .3in}

\caption{Left: Projection function $f(x)$ at distance $x$ from vertical mid-line,  calculated for systems with 9 (blue, top curve), 17 (magenta), 33 (green) and 91 (red, bottom curve) zero-energy modes, in two dimensions.  Inset shows mid-line projection $f(0)$ as a function of the number of particles $N$ (related to the number of zero-energy modes $m$ by $N = m^2$), in two dimensions. Asymptotic power is -1/2.  Right: analogous data from Ref. \cite{Sussman:2016kq}.  Power law in the inset is about -0.4, consistent with -1/2.}
\label{fig:projection}
\end{figure}

To avoid the need to choose a particular basis, it is possible to define a ``projection function" $f_i$ \cite{Wyart2005} 
\begin{equation}
f_i \definedas \frac{1}{m}\sum_p [(\delta x_i^p)^2+(\delta y_i^p)^2],
\end{equation}
where $\delta x_i^p$ is the $x$-displacement of particle $i$ in mode $p$.  Any particle's displacement may be represented as a sum over orthonormal basis vectors.  The quantity $f_i$ is the squared length of the projection of this displacement along a basis spanning the free modes (Ref. \cite{Wyart2005} calls $f_i$ the ``overlap function."  We use ``projection function" instead, in order to avoid confusion with the geometric overlap discussed in Sec. \ref{sec:preparation}). This quantity is basis-independent. If mode $p$ decays with characteristic length $W/(pa)$, then the projection function when there are $m$ modes in a square system centered on $x = 0$ is
\begin{equation}
f(x) = \frac{1}{m}\sum_{p = -(m-1)/2}^{(m-1)/2}\frac{pa}{W\sinh(pa)}~\exp(\frac{2pax}{W}),
\end{equation}
where the normalization ensures $\int_{-W/2}^{W/2}f(x)~dx = 1$, just as $\sum_i f_i = 1$ in Eq. 8.  Figure \ref{fig:projection} shows this projection function for several different values of $m$, using $a = 0.6$ as found empirically in this work. The results are in close agreement with the observations of Sussman et al. \cite{Sussman:2016kq}, with the assumption that $m \approx \sqrt{N}$.

The approach shown in Fig. \ref{fig:projection} can readily be extended to three dimensions.  Here our preliminary findings show behavior of $f(x)$ similar to Fig. \ref{fig:projection}; they require that $N^{2/3} f(0)\goesto 0$ as system size $N\goesto \infinity$.  Our $f(x)$ profiles also resemble the simulation results of Fig. 8 of  Wyart {\sl et al.} \cite{Wyart2005}.  Though these authors interpreted their results as supporting a nonzero $N^{2/3} f(0)$ as $N\goesto \infinity$, it is now acknowledged \cite{Yan:2016fk} that this interpretation is in doubt.

\section{Conclusions and outlook}
\tw{The construction of the most pessimistic basis demonstrates that distinguishable zero-energy modes of motion decay exponentially with depth away from the boundary of a periodic granular system at the jamming transition. This puts a fundamental limit on the amount of information about a perturbation on one side of such a system that can be gained by making measurements on the other side. In particular, the existence of many distinct perturbations, each with a long-ranged response\tw{\cite{Caballero:2005qv,Kolb:2006uq,Shaebani:2008fk}}, does not imply that these responses can be distinguished \tw{via} measurements in a distant region.  \tw{Indeed, in our simulations} the long-ranged component of each response comes from the same, limited set of basis modes that have decay rates comparable to the size of the system.}

It will be very interesting to study systems where this division into local parts is not possible. Such a system can be constructed, even with purely local interactions between particles, simply by changing the boundary conditions. One specific example is a very large system with periodic boundary conditions in both directions, with a small circular hole cut at the center. In such a system, the number of zero-energy modes is determined by conditions at the cut boundary, but the system is constrained by the long-distance periodic boundary conditions. An attempt to replicate the argument made in the discussion of this paper would result in strips with many more zero-energy modes than the system as a whole. It is thus not at all clear that the modes of motion would decay with distance from the circular hole in the same way as those of an elastic material (which exhibit a power-law falloff).

More generally, one could examine the localisation of individual modes and modes in the pessimistic basis that arise on cutting multiple bonds in an extremely large marginal system. For example, if two bonds are cut, the resulting space of zero-energy motion is two-dimensional. Is there a basis of this space in which one of the modes falls off with distance from the cut bonds, and if so, how does the decay length depend on the separation of the cut bonds? Similar questions could be asked for three or more cut bonds. \tw{Experimentally, the elastic bag geometry of Ref. \cite{Brown2010} offers an attractive way of probing transmittivity.  There one may readily perturb individual grains touching the elastic bag and probe the responses of grains elsewhere around the bag.  By perturbing multiple grains simultaneously, one may readily assess the transmittivity of simultaneous signals.  
}

\section{Acknowledgments}
The authors are grateful to Arvind Murugan, Sidney Nagel, Heinrich Jaeger and Tom Lubensky for helpful discussions.  This work was supported in part by the National Science Foundation's MRSEC
Program under Award Number DMR-1420709.  M.P. was supported by a ConocoPhillips fellowship from the American Australian Association.  
\section*{References}
\bibliography{bibliography}{}

\begin{thebibliography}{10}

\bibitem{Silbert2005}
Leonardo Silbert, Andrea Liu, and Sidney Nagel.
\newblock {Vibrations and Diverging Length Scales Near the Unjamming
  Transition}.
\newblock {\em Physical Review Letters}, 95(9):098301, aug 2005.

\bibitem{Duri:2009yq}
A.~Duri, D.~A. Sessoms, V.~Trappe, and L.~Cipelletti.
\newblock Resolving long-range spatial correlations in jammed colloidal systems
  using photon correlation imaging.
\newblock {\em PHYSICAL REVIEW LETTERS}, 102(8):085702, Feb 2009.

\bibitem{vanHecke2010}
M~van Hecke.
\newblock Jamming of soft particles: geometry, mechanics, scaling and
  isostaticity.
\newblock {\em Journal of Physics: Condensed Matter}, 22(3):033101, 2010.

\bibitem{Brown2010}
Eric Brown, Nicholas Rodenberg, John Amend, Annan Mozeika, Erik Steltz,
  Mitchell~R Zakin, Hod Lipson, and Heinrich~M Jaeger.
\newblock Universal robotic gripper based on the jamming of granular material.
\newblock {\em Proceedings of the National Academy of Sciences},
  107(44):18809--18814, 2010.

\bibitem{Wyart2005}
Matthieu Wyart, Leonardo~E Silbert, Sidney~R Nagel, and Thomas~A Witten.
\newblock Effects of compression on the vibrational modes of marginally jammed
  solids.
\newblock {\em Physical Review E}, 72(5):051306, 2005.

\bibitem{Lerner2013}
Edan Lerner, Gustavo D{\"{u}}ring, and Matthieu Wyart.
\newblock {Low-energy non-linear excitations in sphere packings}.
\newblock {\em Soft Matter}, 9(34):8252, 2013.

\bibitem{Muller2015}
Markus M{\"{u}}ller and Matthieu Wyart.
\newblock {Marginal Stability in Structural, Spin, and Electron Glasses}.
\newblock {\em Annual Review of Condensed Matter Physics}, 6(1):177--200, 2015.

\bibitem{LandauElasticity}
Lev~Davidovich Landau and Eugin~M Lifshitz.
\newblock {\em Course of theoretical physics, Theory of elasticity}, chapter~7.
\newblock Pergamon Press Oxford, 1986.

\bibitem{OHern2002}
Corey~S O'Hern, Stephen~A Langer, Andrea~J Liu, and Sidney~R Nagel.
\newblock Random packings of frictionless particles.
\newblock {\em Physical Review Letters}, 88(7):075507, 2002.

\bibitem{OHern2003}
Corey~S O'Hern, Leonardo~E Silbert, Andrea~J Liu, and Sidney~R Nagel.
\newblock Jamming at zero temperature and zero applied stress: The epitome of
  disorder.
\newblock {\em Physical Review E}, 68(1):011306, 2003.

\bibitem{donev2004linear}
Aleksandar Donev, Salvatore Torquato, Frank~H Stillinger, and Robert Connelly.
\newblock A linear programming algorithm to test for jamming in hard-sphere
  packings.
\newblock {\em Journal of Computational Physics}, 197(1):139--166, 2004.

\bibitem{Kapko2009}
V.~Kapko, M.~M.~J. Treacy, M.~F. Thorpe, and S.~D. Guest.
\newblock {On the collapse of locally isostatic networks}.
\newblock {\em Proceedings of the Royal Society A: Mathematical, Physical and
  Engineering Sciences}, 465(2111):3517--3530, 2009.

\bibitem{Matlab}
Mathworks.
\newblock linprog, 2016.

\bibitem{Maxwell1864}
J~Clerk Maxwell.
\newblock On the calculation of the equilibrium and stiffness of frames.
\newblock {\em The London, Edinburgh, and Dublin Philosophical Magazine and
  Journal of Science}, 27(182):294--299, 1864.

\bibitem{Sadd:2014fk}
M.H. Sadd.
\newblock {\em Elasticity: Theory, Applications, and Numerics}, chapter~8.
\newblock Academic Press. Academic Press, 2014.

\bibitem{Mao:2010fk}
X.~M. Mao, N.~Xu, and T.~C. Lubensky.
\newblock Soft modes and elasticity of nearly isostatic lattices: Randomness
  and dissipation.
\newblock {\em PHYSICAL REVIEW LETTERS}, 104(8):085504, Feb 2010.

\bibitem{Lieleg:2011uq}
O.~Lieleg, J.~Kayser, G.~Brambilla, L.~Cipelletti, and A.~R. Bausch.
\newblock Slow dynamics and internal stress relaxation in bundled cytoskeletal
  networks.
\newblock {\em NATURE MATERIALS}, 10(3):236--242, Mar 2011.

\bibitem{Kane2014}
C~L Kane and T~C Lubensky.
\newblock Topological boundary modes in isostatic lattices.
\newblock {\em Nature Physics}, 10(1):39--45, 2014.

\bibitem{Cates1999}
M.~E. Cates, J.~P. Wittmer, J.-P. Bouchaud, and P.~Claudin.
\newblock {Jamming and static stress transmission in granular materials.}
\newblock {\em Chaos (Woodbury, N.Y.)}, 9(3):511--522, 1999.

\bibitem{Sussman:2016kq}
Daniel~M Sussman, Olaf Stenull, and T~C Lubensky.
\newblock Topological boundary modes in jammed matter.
\newblock {\em Soft Matter}, 12:6079--6087, Jun 2016.

\bibitem{Yan:2016fk}
Le~Yan, Eric DeGiuli, and Matthieu Wyart.
\newblock On variational arguments for vibrational modes near jamming.
\newblock {\em EPL (Europhysics Letters)}, 114(2):26003, 2016.

\bibitem{Caballero:2005qv}
G~Caballero, E~Kolb, A~Lindner, J~Lanuza, and E~Cl\'ement.
\newblock Experimental investigation of granular dynamics close to the jamming
  transition.
\newblock {\em Journal of Physics: Condensed Matter}, 17(24):S2503, 2005.

\bibitem{Kolb:2006uq}
Evelyne Kolb, Chay Goldenberg, Shio Inagaki, and Eric Cl\'ement.
\newblock Reorganization of a two-dimensional disordered granular medium due to
  a small local cyclic perturbation.
\newblock {\em Journal of Statistical Mechanics: Theory and Experiment},
  2006(07):P07017, 2006.

\bibitem{Shaebani:2008fk}
M.~Reza Shaebani, Tam\'as Unger, and J\'anos Kert\'esz.
\newblock Unjamming due to local perturbations in granular packings with and
  without gravity.
\newblock {\em Phys. Rev. E}, 78:011308, Jul 2008.

\end{thebibliography}
\bibliographystyle{unsrt}
\end{document}